\documentclass{PoS}
\usepackage{amsmath,amsthm,amsfonts,amssymb}
\usepackage{mathbbol}

\title{Moments of structure functions for $N_f=2$ near the physical point}

\ShortTitle{Moments of structure functions near the physical point}

\author{G.~S.~Bali, \speaker{S.~Collins}, B.~Gl\"a{\ss}le, M.~G\"ockeler, J.~Najjar, R.~R\"odl, A.~Sch\"afer, R.~Schiel, A.~Sternbeck, W.~S\"oldner\\
       Institut f\"ur Theoretische Physik, Universit\"at Regensburg, 93040 Regensburg, Germany\\
        E-mail: \email{sara.collins@physik.uni-regensburg.de}}

\abstract{We report on our on-going study of the lower moments of
  iso-vector polarised and unpolarised structure functions, $g_A$ and
  $\langle x\rangle_{u-d}$, respectively, and the iso-vector scalar
  and tensor charge, for $N_f=2$ non-perturbatively improved clover
  fermions. With pion masses which go down to about $150$~MeV, we
  investigate finite volume effects and excited state contributions.}

\FullConference{31st International Symposium on Lattice Field Theory - LATTICE 2013\\
		July 29 - August 3, 2013\\
		Mainz, Germany}

\begin{document}

\section{Introduction}

In recent years it has become clear that $g_A$ and $\langle
x\rangle_{u-d}$, benchmark quantities for lattice calculations of
nucleon structure, are sensitive to a number of sources of systematic
error - finite volume, non-physical pion mass, excited state
contamination and finite lattice spacing.  Continuing improvements in
computing power, algorithms and analysis techniques mean systematics
can now be investigated and in some cases removed.

\begin{figure}
\begin{center}
\centerline{
\includegraphics[width=0.5\textwidth]{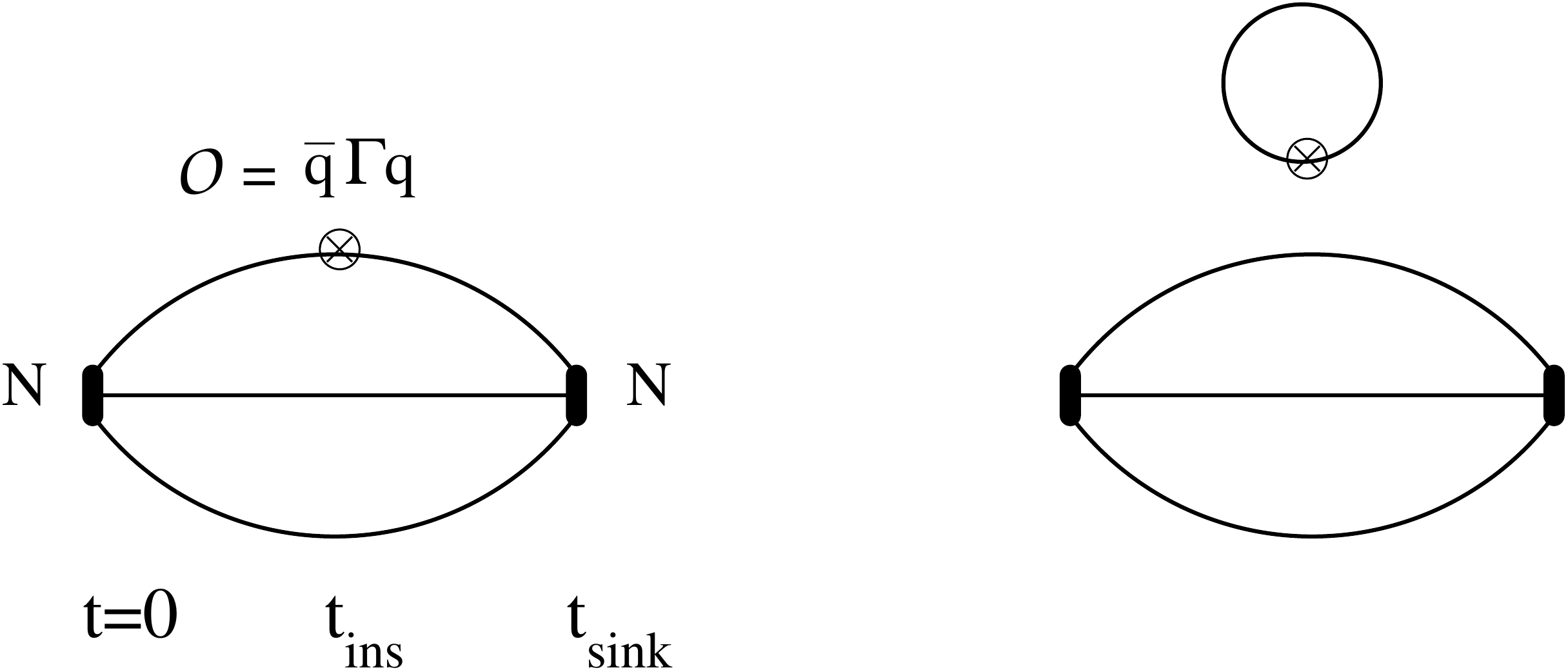}
}
\caption{The connected~(left) and disconnected~(right) contributions
  to nucleon three-point functions for a nucleon source at $t=0$, sink at $t_{\mathrm{sink}}$ and operator insertion at $t_{\mathrm{ins}}$.}\label{conndis}
\end{center}
\end{figure}

In the following we present preliminary results for these quantities
along with the iso-vector scalar and tensor charges, $g_S$ and $g_T$,
respectively, for $N_f=2$ ensembles including different lattice
spacings, pion masses and volumes focusing on studying excited
state effects. The iso-vector combination only requires the calculation of
the connected quark diagram, see Figure~\ref{conndis}. We have also
determined the disconnected contributions to the scalar matrix
element. This and the iso-vector generalised form factors for a range
of operators are detailed in~\cite{Sternbeck}. While many lattice
simulations now include dynamical strange quarks, so far the strange
quark has been found to play a minor role in nucleon structure and
$N_f=2$ simulations are still relevant.

\section{Simulation details}

The results were computed using ensembles generated by QCDSF and the
Regensburg Group, with $N_f=2$ degenerate flavours of dynamical sea
quarks using the non-perturbatively improved clover action at two
lattice spacings and a range of pion masses from $m_{\pi}\sim
490-150$~MeV, see Table~\ref{sim}. Two volumes are available for two
values of $m_{\pi}$, in particular at the near physical point.

The two-point and three-point functions were computed using Wuppertal
smeared sources and sinks with APE smeared gauge links.  For each
ensemble the smearing was optimised to minimize the excited state
contributions to the nucleon two-point function. The connected three-point
functions were generated using the standard sequential propagator
method which involves fixing the sink timeslice~($t_{\mathrm{sink}}$). An
alternative approach using stochastic estimates has been investigated,
see~\cite{Bali:2013gxx}. The values of $t_{\mathrm{sink}}$ were chosen using ensemble
II. For $t_{\mathrm{sink}}=15a\sim 1$~fm on this ensemble no significant excited state
contributions to the quantities $\langle x\rangle_{u-d}$ and $g_A$
were found. This was checked by performing an excited state analysis
with multiple $t_{\mathrm{sink}}$s, described in the next
section. $t_{\mathrm{sink}}=15a$ was then used for all $\beta=5.29$ ensembles
and rescaled for $\beta=5.40$. Multiple measurements were performed on
each configuration. Autocorrelations were investigated by binning the
data with different bin sizes.

\begin{table}
\begin{center}
\begin{tabular}{|c|cccccccc|}\hline
 & $\beta$ & $\kappa$ & Volume  & $N\times M$ & $a/$~fm &  $m_{\pi}/$~MeV & $Lm_\pi$ & $t_{\mathrm{sink}}/a$\\\hline
I &$5.29$    &  0.13620 & $24\times 48$  & $1124\times 2$ & 0.07 &  430  & 3.7 & 15  \\
II  &    & 0.13632&  $32\times 64$  & $2027\times 2~(1)$ &  & 294   & 3.4  & (7,9,11),13,15,17 \\
III  &  & 0.13632&  $40\times 64$  & $2028\times 2$ &  & 289   & 4.2  & 15\\
IV  &  & 0.13640&  $48\times 64$  & $3400\times 2$ &  & 157   & 2.7  & 15\\
V   & & 0.13640&  $64\times 64$  &  $940\times 3$ &  & 150   & 3.5  & 15\\\hline
VI &$5.40$ & 0.13640 & $32\times 64$ & $1170\times 2$ & 0.06 & 491 & 4.8 & 17\\
VII &      & 0.13660 & $48\times 64$ & $2178\times 2$ & & 260 & 3.8 & 17\\\hline
\end{tabular}
\caption{Details of the ensembles used in our analysis including approximate lattice spacings and pion masses and the values of $t_{\mathrm{sink}}$ used for the sequential propagators. The number of configurations~($N$) and measurements made per configuration~($M$) are indicated. For $t_{\mathrm{sink}}/a=7,9$ and $11$, ensemble II, only one measurement per configuration is made. }
\label{sim}
\end{center}
\end{table}

In order to reduce the discretisation effects to $O(a^2)$ , the
operators, as well as the quark action, need to be~(nonperturbatively)
improved. The $O(a)$ improved renormalised operators
have the form~\cite{Capitani:2000xi}
\begin{eqnarray}
& O^{improv} & = Z_O\left[(1+b_Oam_q)O + ac_O O^\prime \right].
\end{eqnarray}
For $g_A$, $O=\bar{q}\gamma_\mu\gamma_5 q$, and the improvement term, $ac_O O^\prime = ac_A\partial_\mu\bar{q}\gamma_5 q$,
does not contribute for forward matrix elements. We use the $Z_O$ factors determined
non-perturbatively~\cite{Gockeler:2010yr,Constantinou:2013ada} and the $b_O$ factors
from~\cite{Sint:1997jx}. Setting $c_O=0$, our values for $g_A$
will have leading $O(a^2)$ effects, while $\langle x\rangle_{u-d}$,
$g_S$ and $g_T$ have $O(a)$.

\section{Excited state fits}

Excited state contributions to nucleon structure have been investigated
by a number of groups recently, see, for
example,~\cite{Dinter:2011sg,Owen:2012ts,Capitani:2012gj,Green:2012ud,Bhattacharya:2013ehc}. Our
analysis on ensemble II is similar to that performed in
reference~\cite{Bhattacharya:2013ehc}. We fitted the
two-point~($C_{2pt}$) and three-point~($C_{3pt}$) functions at
multiple $t_{\mathrm{sink}}$s simultaneously using the functional forms:
\begin{eqnarray}
C_{2pt}(t_{\mathrm{sink}}) &= & \sum_{\vec{x}}\langle {\cal N}(\vec{x},t_{\mathrm{sink}})\overline{{\cal N}}(\vec{0},0)\rangle = |Z_0|^2e^{-m_0 t_{\mathrm{sink}}}+ |Z_1|^2e^{-m_1 t_{\mathrm{sink}}}+\ldots\\
& = &  e^{-m_0 t_{\mathrm{sink}}} [|Z_0|^2+ |Z_1|^2e^{-\Delta m t_{\mathrm{sink}}}+\ldots]\label{twofit}\\
C_{3pt}(t_{\mathrm{sink}},t_{\mathrm{ins}}) & = &\sum_{\vec{x},\vec{y}}\langle {\cal N}(\vec{x},t_{\mathrm{sink}}) O(\vec{y},t_{\mathrm{ins}})\overline{{\cal N}}(\vec{0},0)\rangle\\
 & =&  |Z_0|^2\langle N_0|O|N_0\rangle e^{-m_0 t_{\mathrm{sink}}}+ Z_1^*Z_0 \langle N_1|O|N_0\rangle e^{-m_0 t_{\mathrm{ins}}}e^{-m_1(t_{\mathrm{sink}}-t_{\mathrm{ins}})}  \nonumber\\
& & + Z_0^*Z_1 \langle N_0|O|N_1\rangle e^{-m_1 t_{\mathrm{ins}}}e^{-m_0(t_{\mathrm{sink}}-t_{\mathrm{ins}})} + |Z_1|^2\langle N_1|O|N_1\rangle e^{-m_1 t_{\mathrm{sink}}}+\ldots\\
& =& |Z_0|^2 e^{-m_0t_{\mathrm{sink}}}\left(B_0 + B_1 [ e^{-\Delta m (t_{\mathrm{sink}}-t_{\mathrm{ins}})} + e^{-\Delta m t_{\mathrm{ins}}}] + B_2 e^{-\Delta m t_{\mathrm{sink}}}\right)+\ldots\label{threefit}
\end{eqnarray}
where ${\cal N}$ destroys the nucleon, $O$ is the current insertion
and $Z_i=\langle 0|{\cal N}|N_i\rangle$. $|N_0\rangle$ and
$|N_1\rangle$ represent the nucleon ground and first excited state,
respectively. For the cases considered here, it is sufficient to consider zero initial and final momentum in order to extract forward matrix
elements.

The matrix element of interest is given by $B_0= \langle
N_0|O|N_0\rangle$ while $B_1 \propto \langle N_1|O|N_0\rangle$ gives
the transition matrix element from the ground to the first excited state
and $B_2 \propto \langle N_1|O|N_1\rangle$ the first excited state
matrix element. These fits can be compared to the traditional approach
of fitting the ratio of three-point to two-point functions to a constant:
\begin{equation}
\frac{C_{3pt}(t_{\mathrm{sink}},t_{\mathrm{ins}}) }{C_{2pt}(t_{\mathrm{sink}})} = B_0 + \ldots \label{constfit}
\end{equation}

\begin{figure}
\begin{center}
\centerline{
\includegraphics[width=0.5\textwidth]{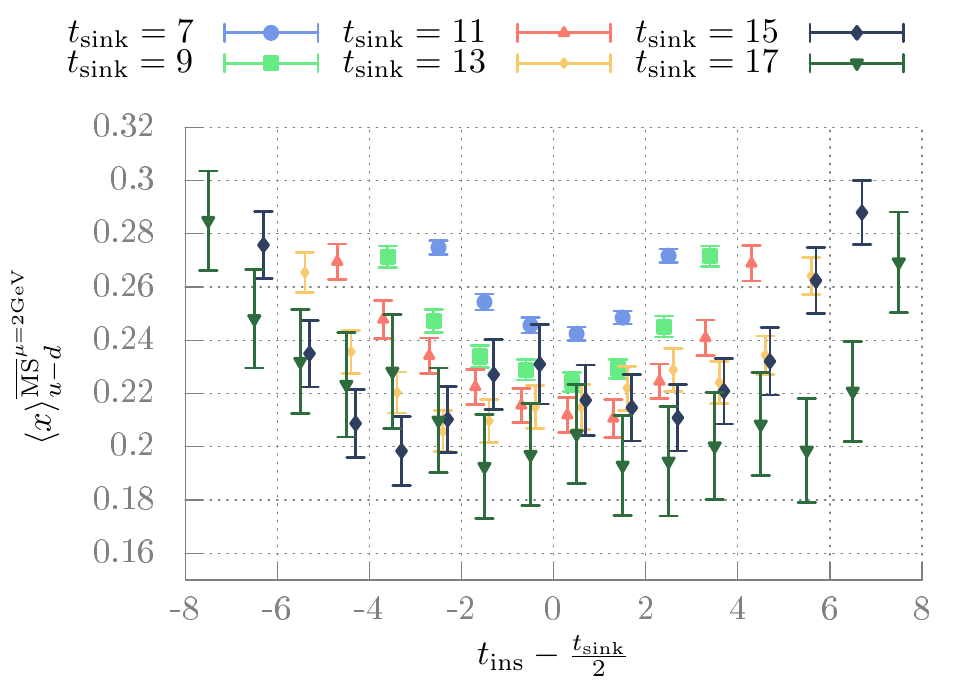}
\includegraphics[width=0.5\textwidth]{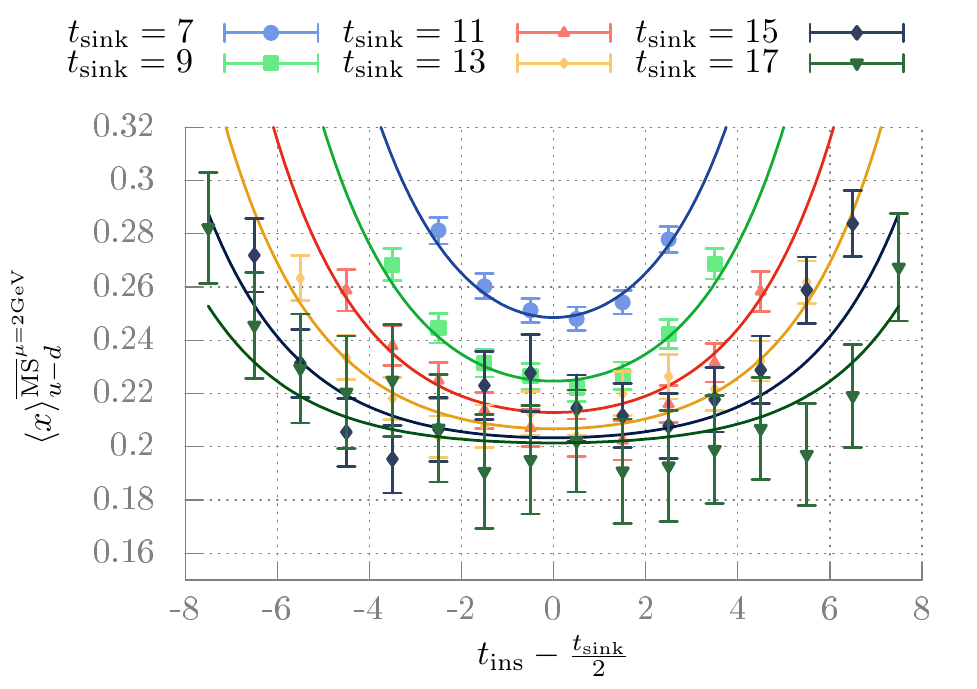}
}
\caption{(Left) The raw ratio $(C_{3pt}/C_{2pt})\cdot\mathrm{factor}$ for
  ensemble II for different $t_{\mathrm{sink}}$s, where the factor includes the renormalisation
 and improvement constants and mass term necessary to convert to $\langle x\rangle_{u-d}^{\overline{\mathrm{MS}}}$ at ${\mu=2\,\mathrm{GeV}}$.
  (Right) The raw results for $C_{3pt}$, ensemble II, divided by the fit result for $|Z_0|^2e^{-m_0t_{\mathrm{sink}}}$, compared to the corresponding fit function divided by the same quantity: $B_0 + B_1 [ e^{-\Delta m (t_{\mathrm{sink}}-t_{\mathrm{ins}})} + e^{-\Delta m t_{\mathrm{ins}}}] + B_2 e^{-\Delta m t_{\mathrm{sink}}}$. The same factor is included as for the figure on the left in order to convert
 to $\langle x\rangle_{u-d}^{\overline{\mathrm{MS}}}$ at ${\mu=2\,\mathrm{GeV}}$. }
\label{fitallx}
\end{center}
\end{figure}

\begin{figure}
\begin{center}
\centerline{
\includegraphics[width=0.5\textwidth]{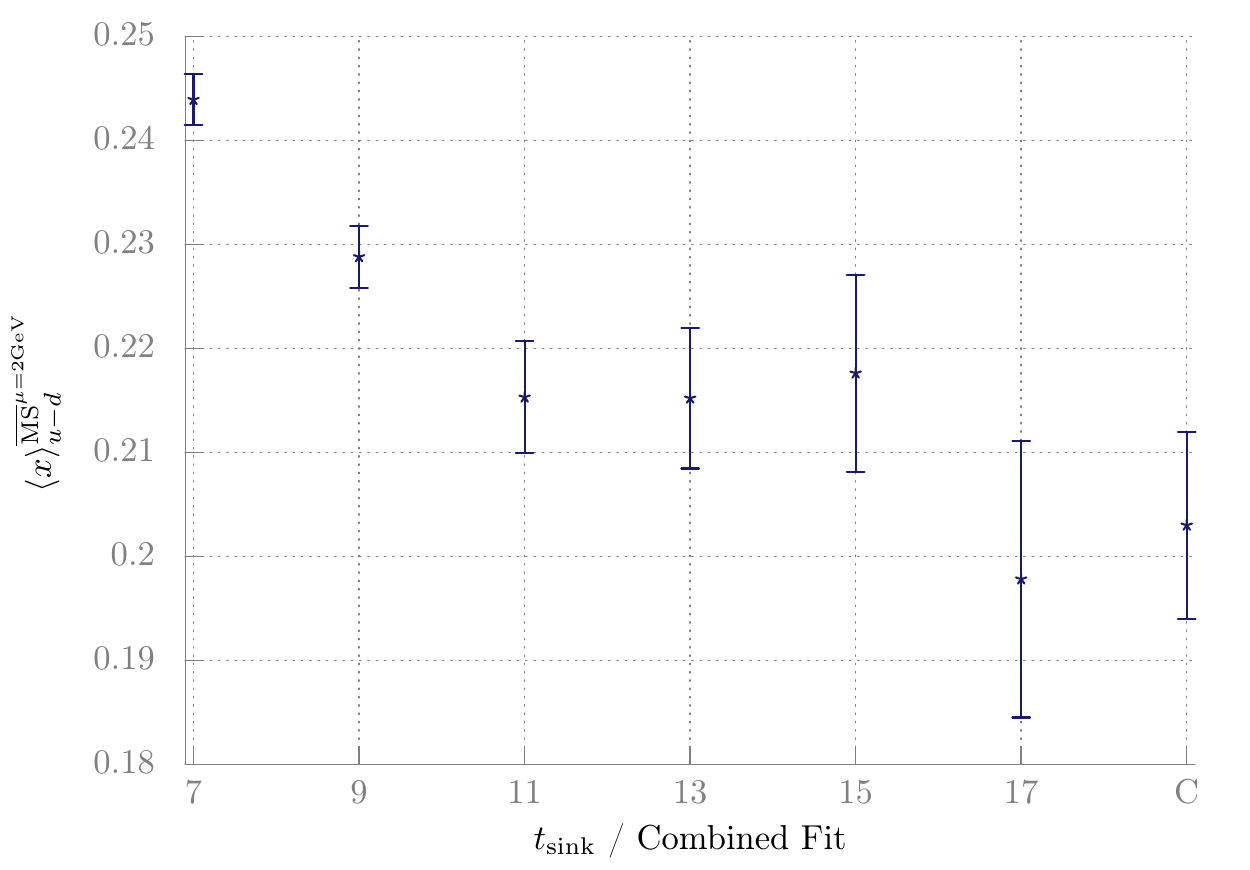}
\includegraphics[width=0.55\textwidth]{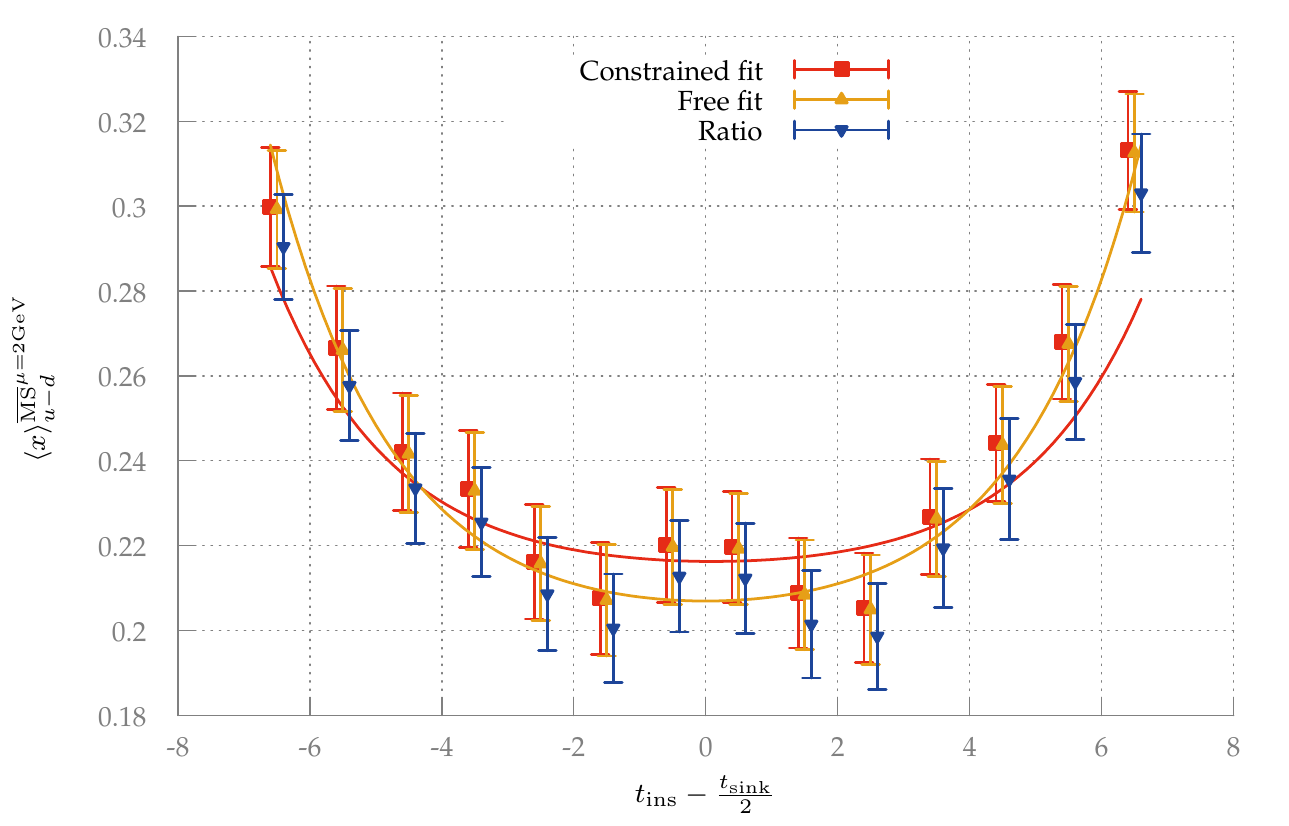}
}
\caption{(Left) For ensemble II, $\langle
  x\rangle_{u-d}^{\overline{\mathrm{MS}}^{\mu=2\mathrm{GeV}}}$
  extracted from constant fits to the ratio
  $C_{3pt}/C_{2pt}$~(Eq.~(\protect\ref{constfit})) for different
  $t_{\mathrm{sink}}$ and extracted from the combined fit to $C_{3pt}$ and
  $C_{2pt}$, for all $t_{\mathrm{sink}}$s, including the first excited state,
    denoted by $C$. (Right) For ensemble IV, the raw ratio
    $C_{3pt}/C_{2pt}\cdot\mathrm{factor}$ is shown~(red triangles). The
    two lines indicate fits (i) constraining the excited state
    parameters using the results from ensemble II~(blue line) and (ii)
    leaving $B_1$ free and $B_2=0$~(green line).  For each of these
    fits the raw results for $C_{3pt}\cdot\mathrm{factor}$, divided by
    $|Z_0|^2e^{-m_0t_{\mathrm{sink}}}$, is shown.}
\label{secondfits}
\end{center}
\end{figure}

We illustrate the results for the example of $\langle x\rangle_{u-d}$,
which we found to have significant excited state contributions at
small $t_{\mathrm{sink}}$. Figure~\ref{fitallx} displays the raw
results for the six $t_{\mathrm{sink}}$ values. Agreement is found for
$t_{\mathrm{sink}}\ge 11a$ for
$t_{\mathrm{ins}}-t_{\mathrm{sink}}/2\approx 0$. For
$t_{\mathrm{sink}}\ge 13$a, the plateau extends for several timeslices,
however, as expected, the statistical errors also increase. Correlated
fits to all three-point and two-point functions using
Eqs.~(\ref{twofit}) and~(\ref{threefit})~(truncating after the first
excited state) produced good $\chi^2/d.o.f \approx 1$ and were stable
against changes in the fitting ranges to $C_{3pt}$ and $C_{2pt}$.
An example of one of the combined fits is shown in Figure~\ref{fitallx}. For
the smearing we have used, the ground and first excited state are the
dominant contributions for $t_{\mathrm{sink}}$ as small as $7a$.  In
Figure~\ref{secondfits}, a result of the combined fit is compared to
the values obtained fitting the ratio $C_{3pt}/C_{2pt}$ to a constant
for different $t_{\mathrm{sink}}$; consistency is found between the two methods
for $t_{\mathrm{sink}}\ge 11a$. From this analysis we find our optimised
smearing is sufficient to extract the ground state matrix element
using a single $t_{\mathrm{sink}}\ge 11a$ and we take the conservative choice of
$15a$. We remark that with a less optimised smearing we found a   $t_{\mathrm{sink}}>15a$
insufficient.

For the other ensembles, where we only have one $t_{\mathrm{sink}}$,
we check the size of the excited state contamination by performing
fits using Eqs.~\ref{twofit} and~\ref{threefit} and the
parameters proportional to the excited state matrix elements, 
$B_1$ and $B_2$, extracted using ensemble II~(which we assume to be
slowly varying with the quark mass and lattice
spacing). Figure~\ref{secondfits} shows such a constrained fit for
ensemble IV. The value extracted for the ground state is consistent
with a fit using the same functional form, but with $B_1$ as a free
parameter and $B_2=0$~\footnote{This parameter cannot be determined
  from fits to a single $t_{\mathrm{sink}}$.}. It is also consistent
with the result of a constant fit to $C_{3pt}/C_{2pt}$.

The excited state fitting analysis was similarly successful for the
other quantities of interest. We note that for $g_A$, significant
excited state contributions were found for $t_{\mathrm{sink}}=7$ and $9$,
however, these contributions cancelled in the ratio $C_{3pt}/C_{2pt}$,
giving a plateau for the ratio, even for the smallest $t_{\mathrm{sink}}$. For
$g_S$, the ground state was dominant for all $t_{\mathrm{sink}}$ for the
iso-vector combination in the ratio $C_{3pt}/C_{2pt}$. This appears to
be due to a cancellation in the $u-d$ combination since for the
iso-scalar matrix elements, not presented in this work, there were
significant excited state contributions. The tensor charge analysis
was similar to that for $\langle x\rangle_{u-d}$.

\section{Results}

\begin{figure}
\begin{center}
\centerline{
\includegraphics[width=0.5\textwidth,clip]{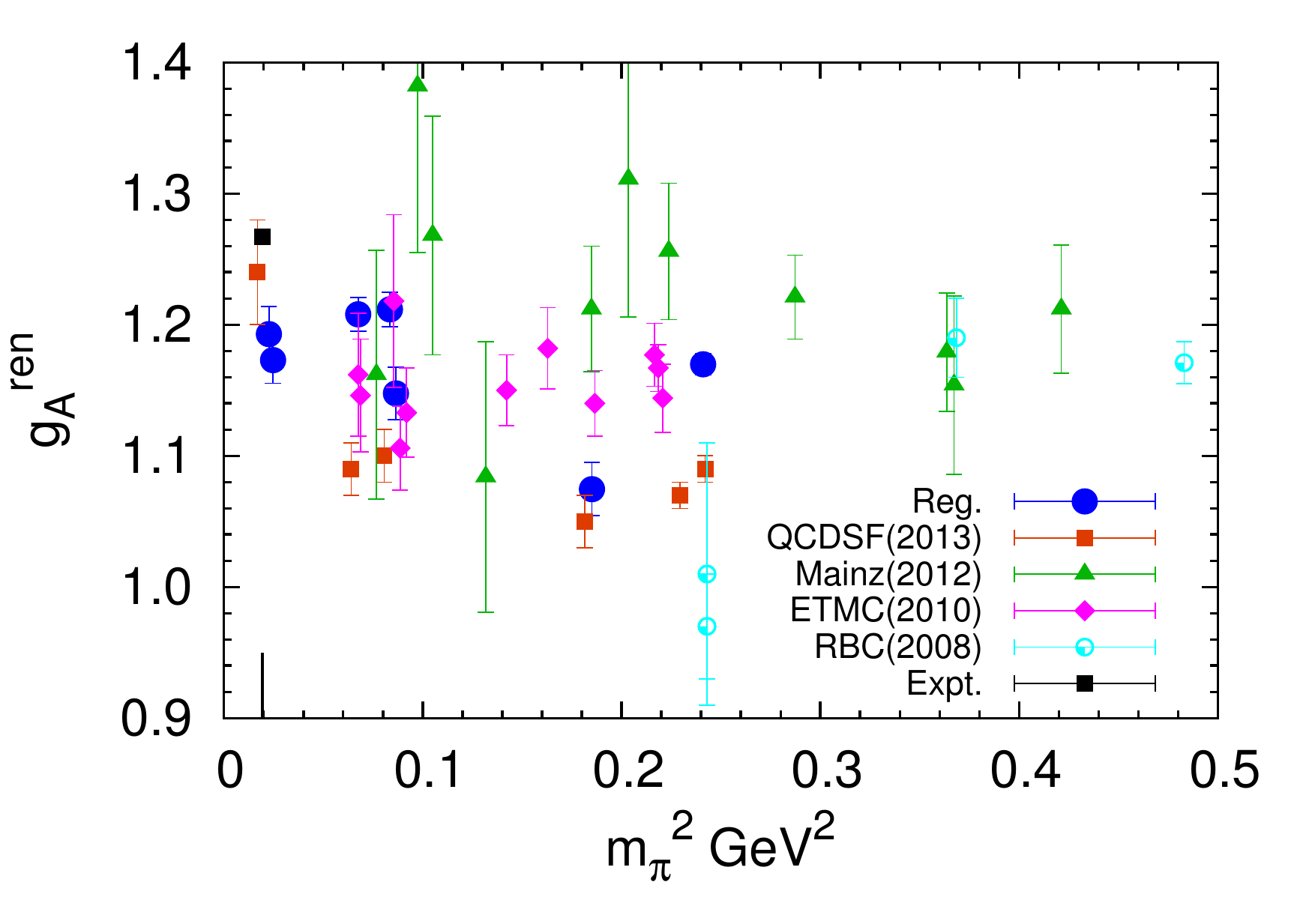}
\includegraphics[width=0.5\textwidth,clip]{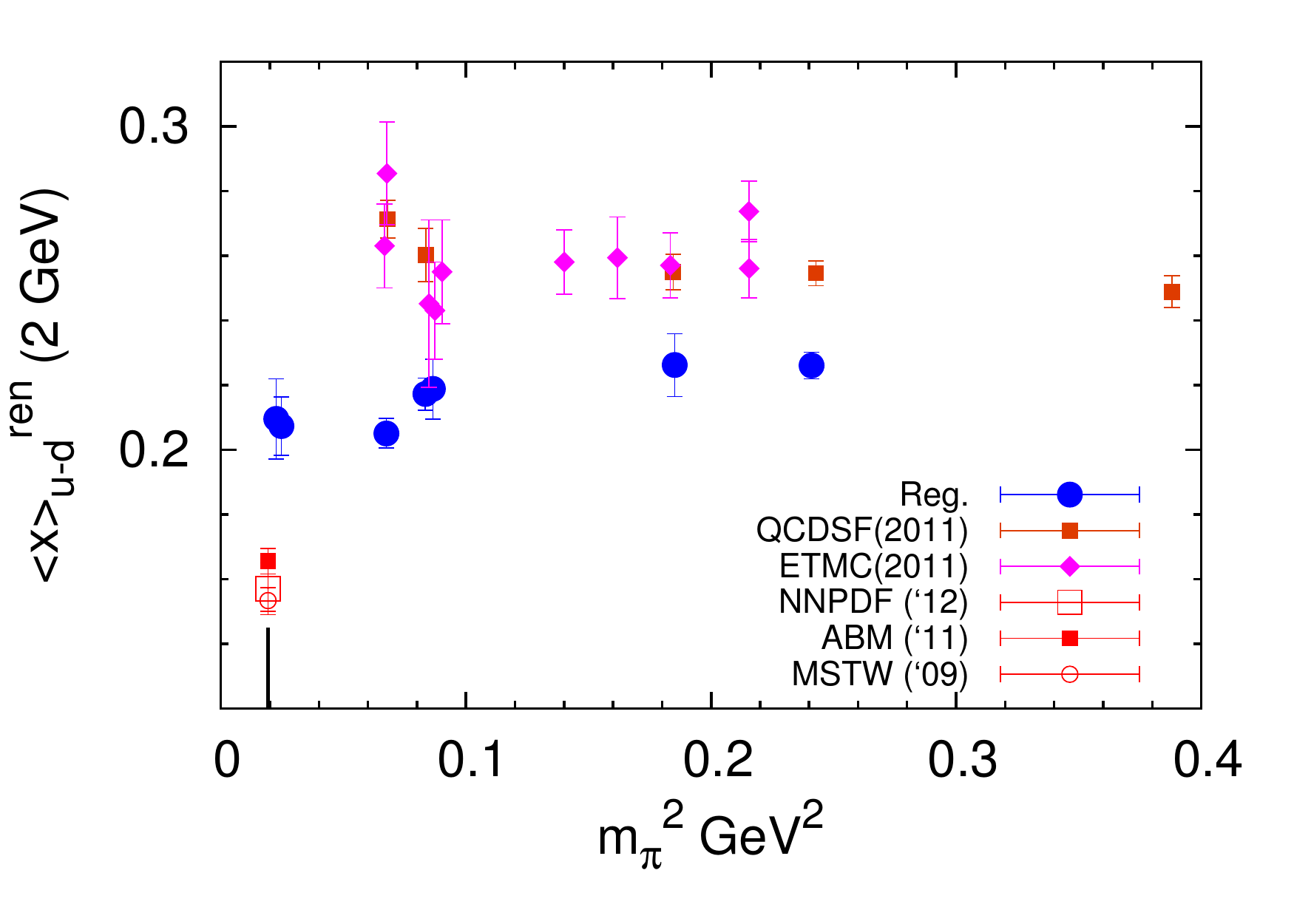}
}
\centerline{
\includegraphics[width=0.5\textwidth,clip]{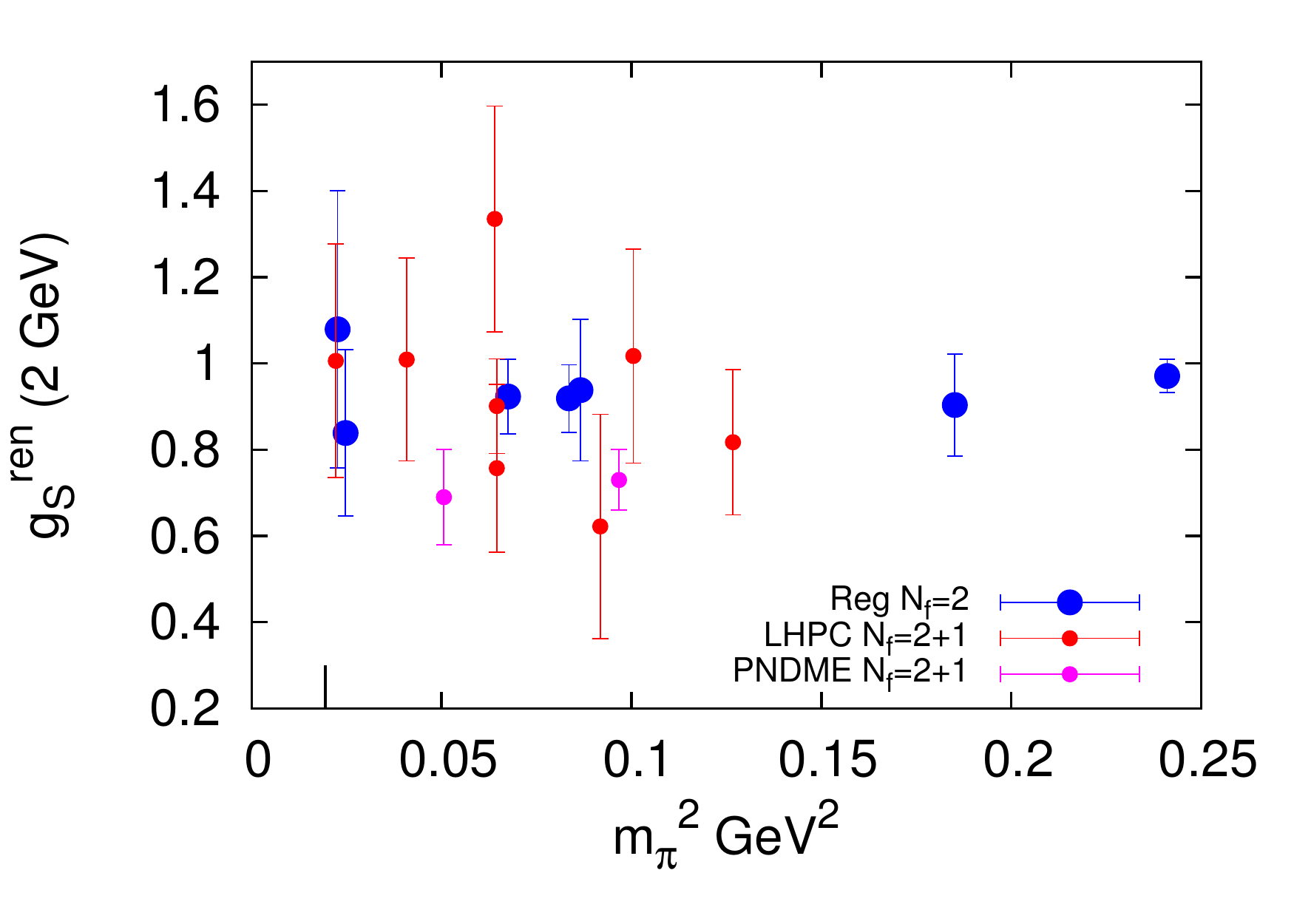}
\includegraphics[width=0.5\textwidth,clip]{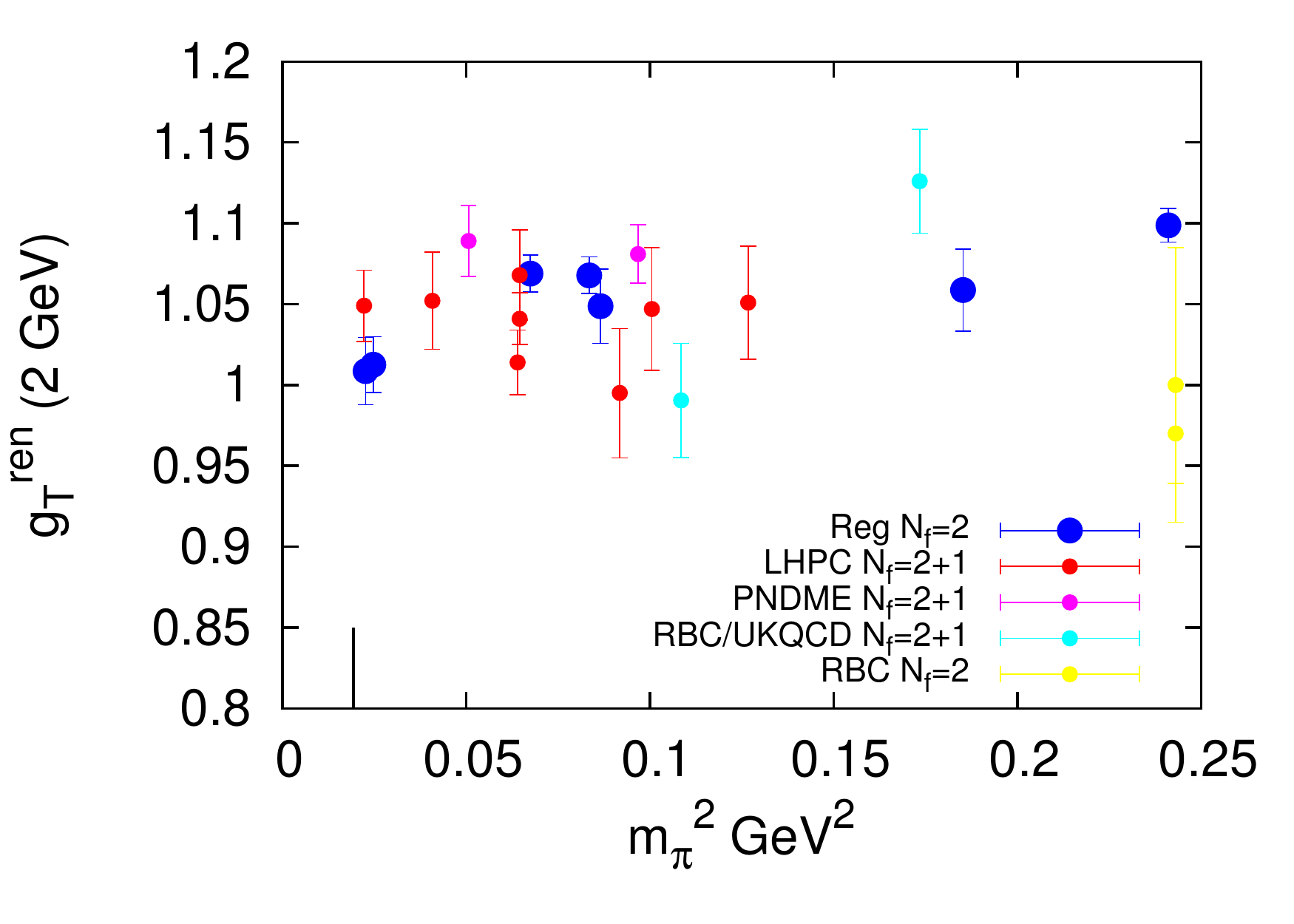}
}
\caption{Results obtained from all ensembles as a function of
  $m_{\pi}^2$~(blue circles) compared to previous works. The physical
  point is indicated by a vertical line. For $g_A$ and $\langle
  x\rangle_{u-d}$ only $N_f=2$ determinations are shown. For the
  former, the experimental result is shown as a black square, while
  for the latter, the expectations from PDF parameterisations of the
  NNPDF, ABM and MSTW groups are shown. }
\label{ourresults}
\end{center}
\end{figure}

In the following we present results obtained from a constant fit to
the ratio $C_{3pt}/C_{2pt}$, for a single $t_{\mathrm{sink}}$.
Figure~\ref{ourresults} shows our results for $g_A$, $\langle
x\rangle_{u-d}$, $g_T$ and $g_S$ from all ensembles as a function of
$m_{\pi}^2$. Recent work from other groups is indicated, where for
$g_A$ and $\langle x \rangle_{u-d}$, due to the large number of
previous determinations, we only compare with other $N_f=2$
calculations. Recent $N_f=2+1$ and $2+1+1$ results are reviewed
in~\cite{Syritsyn}.  For the scalar and tensor charge most other results
are for $N_f=2+1$ and $2+1+1$.

Considering $g_A$ first, one can see in Figure~\ref{ourresults} that
there is a significant dependence of our results on the volume and
possibly $m_\pi$ and $a$. Consistency is found with the values from
the Mainz Group~\cite{Capitani:2012gj} and also
ETMC~\cite{Alexandrou:2010hf}. The discrepancy with
reference~\cite{Horsley:2013ayv}, which uses some of the same
configurations but in some cases different smearing and $t_{\mathrm{sink}}$-values, is mostly likely due to excited state contamination.
The near physical point in that
study is computed on a subset of the statistics~(same smearing and
$t_{\mathrm{sink}}$) for ensemble IV. For the higher statistics used in our
work, $g_A$ is not consistent with the experimental value at this
small volume~($Lm_\pi=2.7$). The extent of finite volume effects at
the near physical point will become clearer once we achieve full
statistics on the larger volume, ensemble V~($Lm_\pi=3.5$).

The effect of excited state contributions can also be seen in the
results for $\langle x\rangle_{u-d}$. Our values lie significantly
below earlier results of QCDSF~\cite{Sternbeck:2012rw,Pleiter:2011gw} and ETMC~\cite{Alexandrou:2011nr}. Such effects have
been seen in previous
works~\cite{Dinter:2011sg,Owen:2012ts,Capitani:2012gj,Green:2012ud,Bhattacharya:2013ehc}. The
two volumes at $\beta=5.29$ and $m_\pi\sim 290$~MeV, and similarly at
the near physical point, indicate that finite volume effects are not
significant for this quantity. Similarly, the pion mass dependence
seems to be mild. The remaining discrepancy with the predictions from
PDF parameterisations may be due to lattice spacing effects.

The scalar coupling suffers from larger statistical errors than the
other quantities considered here. Within the large error, there is no
significant dependence on $m_\pi$, volume or lattice spacing. Our
results are consistent with other recent determinations. A similar
picture is found for the tensor charge, although the statistical
errors are smaller in this case.

\section{Outlook}

Control of excited state contributions and simulation at near physical
pion masses are first steps towards a precise determination of $g_A$
and $\langle x\rangle_{u-d}$. For $g_A$, a careful volume extrapolation
is needed, while in both cases the continuum limit needs to be
studied. We are extending our simulations with this aim.

\section{Acknowledgements}
This work is supported by the EU ITN STRONGnet and the DFG SFB/TRR
55. Computations were performed on SuperMUC of the Leibniz Computing
Center, the Regensburg iDataCool cluster and the SFB/TR55 QPACE
supercomputers.  The Chroma software suite~\cite{Edwards:2004sx} was
used extensively in this work along with the domain decomposition
solver implementation of~\cite{luscherweb}.


\begin{thebibliography}{10}

\bibitem{Sternbeck}
A.~Sternbeck {\em et~al.},
\newblock PoS {\bf Lattice2013}, 291 (2013).

\bibitem{Bali:2013gxx}
G.~S. Bali {\em et~al.},
\newblock (2013), arXiv:1311.1718.

\bibitem{Capitani:2000xi}
S.~Capitani {\em et~al.},
\newblock Nucl.Phys. {\bf B593}, 183 (2001), arXiv:hep-lat/0007004.

\bibitem{Gockeler:2010yr}
M.~G{\"o}ckeler {\em et~al.},
\newblock Phys.Rev. {\bf D82}, 114511 (2010), arXiv:1003.5756.

\bibitem{Constantinou:2013ada}
M.~Constantinou {\em et~al.},
\newblock Phys.Rev. {\bf D87}, 096019 (2013), arXiv:1303.6776.

\bibitem{Sint:1997jx}
S.~Sint and P.~Weisz,
\newblock Nucl.Phys. {\bf B502}, 251 (1997), arXiv:hep-lat/9704001.

\bibitem{Dinter:2011sg}
S.~Dinter {\em et~al.},
\newblock Phys.Lett. {\bf B704}, 89 (2011), arXiv:1108.1076.

\bibitem{Owen:2012ts}
B.~J. Owen {\em et~al.},
\newblock Phys.Lett. {\bf B723}, 217 (2013), arXiv:1212.4668.

\bibitem{Capitani:2012gj}
S.~Capitani {\em et~al.},
\newblock Phys.Rev. {\bf D86}, 074502 (2012), arXiv:1205.0180.

\bibitem{Green:2012ud}
J.~Green {\em et~al.},
\newblock (2012), arXiv:1209.1687.

\bibitem{Bhattacharya:2013ehc}
T.~Bhattacharya, S.~D. Cohen, R.~Gupta, A.~Joseph, and H.-W. Lin,
\newblock (2013), arXiv:1306.5435.

\bibitem{Syritsyn}
S.~Syritsyn,
\newblock PoS {\bf Lattice2013}, 009 (2013).

\bibitem{Alexandrou:2010hf}
ETM Collaboration, C.~Alexandrou {\em et~al.},
\newblock Phys.Rev. {\bf D83}, 045010 (2011), arXiv:1012.0857.

\bibitem{Horsley:2013ayv}
R.~Horsley {\em et~al.},
\newblock (2013), arXiv:1302.2233.

\bibitem{Sternbeck:2012rw}
A.~Sternbeck {\em et~al.},
\newblock PoS {\bf LATTICE2011}, 177 (2011), arXiv:1203.6579.

\bibitem{Pleiter:2011gw}
QCDSF/UKQCD Collaboration, D.~Pleiter {\em et~al.},
\newblock PoS {\bf LATTICE2010}, 153 (2010), arXiv:1101.2326.

\bibitem{Alexandrou:2011nr}
C.~Alexandrou {\em et~al.},
\newblock Phys.Rev. {\bf D83}, 114513 (2011), arXiv:1104.1600.

\bibitem{Edwards:2004sx}
SciDAC, LHPC, UKQCD, R.~G. Edwards and B.~Jo\'o,
\newblock Nucl.Phys.Proc.Suppl. {\bf 140}, 832 (2005), arXiv:hep-lat/0409003.

\bibitem{luscherweb}
http://luscher.web.cern.ch/luscher/DD HMC/index.html.

\end{thebibliography}
\end{document}